\documentclass[10pt]{wlscirep}

\usepackage{filecontents}
\usepackage[utf8]{inputenc}
\usepackage[T1]{fontenc}
\usepackage[english]{babel}
\usepackage{graphicx}
\usepackage{bm}
\usepackage{url}
\usepackage{amsmath}
\usepackage{amssymb}
\usepackage{hyperref}
\hypersetup{colorlinks=true,linkcolor=blue,citecolor=blue,urlcolor=blue}

\title{Ptychography of pure quantum states}

\author[*]{Mário Foganholi Fernandes}
\author[+]{Leonardo Neves}
\affil[ ]{Departamento de Física, Universidade Federal de  Minas Gerais, Belo Horizonte, MG 31270-901, Brazil}

\affil[*]{fernandes-mario@ufmg.br}
\affil[+]{lneves@fisica.ufmg.br}

\begin{abstract}
\sffamily\bf
  Ptychography is an imaging technique in which a localized illumination  scans overlapping regions of an object and generates a set of diffraction intensities used to computationally reconstruct its complex-valued transmission function. We propose a quantum analogue of this technique designed to reconstruct $d$-dimensional pure states. A set of $n$ rank-$r$ projectors ``scans'' overlapping parts of an input state and the moduli of the $d$ Fourier amplitudes of each part are measured. These $nd$ outcomes are fed into an iterative phase retrieval algorithm that estimates the state. Using $d$ up to 100 and $r$ around $d/2$, we performed numerical simulations for single systems in an economic ($n=4$) and a costly ($n=d$) scenario, as well as for multiqubit systems ($n=6\log d$). This numeric study included realistic amounts of depolarization and poissonian noise, and all scenarios yielded, in general, reconstructions with infidelities below $10^{-2}$. The method is shown, therefore, to be resilient to noise and, for any $d$, requires a simple and fast postprocessing algorithm. We show that the algorithm is equivalent to an alternating gradient search, which ensures that it does not suffer from local-minima stagnation. Unlike traditional approaches to state reconstruction, the ptychographic scheme uses a single measurement basis; the diversity and redundancy in the measured data---key for its success---are provided by the overlapping projections. We illustrate the simplicity of this scheme with the paradigmatic multiport interferometer.
\end{abstract}

\begin{document}

\maketitle
\thispagestyle{empty}

Assessing the state of a quantum system is a task of fundamental importance in practical implementations of quantum technologies: one might be interested, for example, in knowing if a certain setup is implementing an intended state, or what is the resulting state after a given evolution. Besides, knowledge of the state allows one to calculate any possible measurement outcome, any quantity relevant for quantum information processing and determine the subsequent evolution of the system. In a standard approach, this task is carried out by making  projective measurements (in appropriate bases) on identically prepared quantum systems, estimating the outcome probabilities, and feeding them into some postprocessing algorithm that will deliver a physical state compatible with the data set \cite{Wootters89,James01}. This process, known as quantum state tomography, has become an integral part of the quantum information toolbox \cite{James01,Thew02,Gross10,Adamson10,Toth10,Maciel11,Lima11}.

The complexity of quantum tomography increases with the state-space dimension, $d$, as the required number of measurement bases (or unitary operations on the system) scales with $d$, at least \cite{Haffner05,Klimov08,Kaznady09}. However, under prior information the process is simplified. For example, if an unknown state is known to be pure, four \cite{Goyeneche08,Goyeneche15,Stefano17} or five \cite{Goyeneche15,Carmeli16} measurement bases and simple postprocessing suffice for determining it on any finite dimension. Yet, to implement the measurements in a variety of bases (or, equivalently, to implement various unitary operations) may not be straightforward in all \text{experiments.}

In this work, we introduce a method for pure state reconstruction that, unlike a typical tomography, uses a single basis in which $n$ partially overlapping parts of the unknown state are measured. Additionally, it employs a simple and fast iterative phase retrieval algorithm for postprocessing. The method is based on \emph{ptychography} \cite{Hoppe69,Rodenburg08,Faulkner04}, a powerful coherent diffractive imaging (CDI) technique with applications in optical \cite{Rodenburg07,Rodenburg07-2,Thibault08} and electron \cite{Humphry12,Jiang18} microscopy, biological \cite{Marrison13} and nonlinear \cite{Odstrcil16} imaging, among others\cite{Maiden11,Shi13,McDermott17}.

\begin{figure*}[tb]
\centerline{\includegraphics[width=1.0\textwidth]{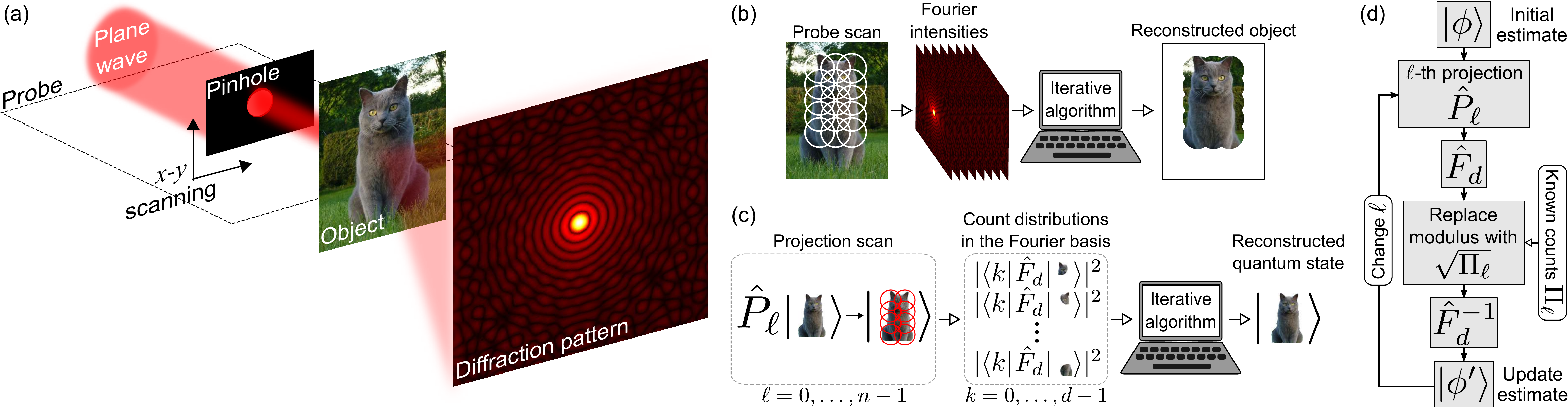}}
\caption{\label{fig:ptico_clas} \textbf{Classical and quantum ptychography.} (a) Schematic of a typical ptychography setup. (b) Ptychographic imaging process. (c) Quantum state ptychography. (d) PIE diagram. See text for details. }
\end{figure*}

A typical setup for ptychography is sketched in Fig.~\ref{fig:ptico_clas}(a): a plane wave filtered by a pinhole creates a localized illumination probe on the object to be imaged; in the far field, one measures the intensity of the generated diffraction pattern (Fourier intensity). The ptychographic CDI process is carried out by scanning the probe over partially overlapping parts of the object and recording the corresponding Fourier intensities.  In the simplest case, where both the illumination probe and its positioning are accurately known {\it a priori}, this data set and the probe information are fed into an iterative algorithm, called ptychographic iterative engine (PIE) \cite{Rodenburg04,Faulkner05}. Starting with a random or uniform estimate for the complex-valued object transmission function, the PIE will iteratively update it by imposing the measured intensities. The implicit phase corrections, resulting from moduli imposition in the conjugate domain, together with the diversity and redundancy in the data, attained by the multiple overlapping illuminations, make the initial estimate converge to the object function. Figure~\ref{fig:ptico_clas}(b) outlines the process.

In the proposed quantum analogue of ptychographic CDI, sketched in Fig.~\ref{fig:ptico_clas}(c), a set of $n$ projectors ``scans'' overlapping parts of an input pure state and the moduli of the $d$ Fourier amplitudes of each part are measured. These $nd$ outcomes are fed into a PIE-based algorithm which estimates the state. For $d$ up to 100, we simulated numerically the ptychographic reconstruction of single and multipartite states for different sets of projectors and, considering realistic noisy scenarios, obtained successful reconstructions in all cases. We also demonstrated the equivalence between our reconstruction algorithm and an alternating gradient search, which ensures that it will not stagnate at local minima. It is important to underline that quantum ptychography is a method to accomplish the task of quantum tomography for pure states. However, unlike most tomographic methods which use measurements in several bases, the ptychographic scheme uses only one basis and several projectors of rank greater than one. From the experimental point of view, this is advantageous as it requires much simpler settings in the setup. We illustrate this point through the paradigmatic multiport interferometer\cite{Reck94}. From the theoretical perspective, the main advantage of quantum ptychography is that it uses a phase-retrieval-like algorithm for postprocessing, which is much simpler than maximum-likelihood\cite{Kaznady09} or semi-definite programming methods\cite{Maciel11}, commonly employed in usual tomographic methods. It would certainly be possible to use these in quantum ptychography, though.

\section*{Quantum state ptychography}

\subsection*{The protocol}

An arbitrary pure quantum state in a $d$-dimensional Hilbert space $\mathcal{H}_d$ may be written in the computational basis $\{|k\rangle\}_{k=0}^{d-1}$ as
\begin{equation}
  |\psi\rangle=\sum_{k=0}^{d-1}c_k|k\rangle,
\end{equation}
where $\sum_k|c_k|^2=1$. This is the object we want to reconstruct by determining the $d$ complex coefficients $\{c_k\}_{k=0}^{d-1}$, which completely specify it. In our ptychographic approach, the role of the localized, scanning, and partially overlapping illumination probe will be played by a set of $n$ projectors $\{\hat{P}_\ell\}_{\ell=0}^{n-1}$ onto $r$-dimensional subspaces ($1<r<d$) of $\mathcal{H}_d$. For this set of rank-$r$ probe projectors we impose two conditions: (i) each projector in the set must have a partial overlap with at least one other partner, i.e., for any $\hat{P}_\ell$ there exists a $\hat{P}_{\ell'}$ such that $0<\mathcal{O}\equiv{\rm Tr}(\hat{P}_\ell\hat{P}_{\ell'})/r<1$; (ii) all levels in $\mathcal{H}_d$ must be addressed at least once. Some examples of such sets are given later on, but until then, the discussion is general and applies to any set of probe projectors.

Given an ensemble of $d$-dimensional quantum systems described by the state $|\psi\rangle$, the ptychographic measurements on this state proceeds as follows: we first apply the $\ell$-th probe projection on the input ensemble, generating an output sub-ensemble described by the (unnormalized) state
\begin{equation}
  |\psi_\ell\rangle=\hat{P}_\ell|\psi\rangle.
\end{equation}
Next, we apply a quantum Fourier transform (QFT) on this output, obtaining
\begin{equation}
|\widetilde{\psi}_\ell\rangle=\hat{F}_d|\psi_\ell\rangle=\sum_{k=0}^{d-1}\tilde{c}_{k\ell}|k\rangle,
\end{equation}
where $\hat{F}_d$ is the QFT acting on $\mathcal{H}_d$ and $\{\tilde{c}_{k\ell}\}_{k=0}^{d-1}$ is the set of Fourier transformed amplitudes of $|\psi_\ell\rangle$. Finally, we perform a projective measurement in the computational basis. This procedure is repeated for each $\hat{P}_\ell$ and gives us a set of $n$ count distributions $\bm{\{}\Pi_\ell=\{\mathcal{N}|\tilde{c}_{k\ell}|^2\}_{k=0}^{d-1}\bm{\}}_{\ell=0}^{n-1}$, where $\mathcal{N}$ is a constant dependent on the particle flux (before projections) and detector efficiencies. These distributions form our ptychographic data set as $\{\sqrt{\Pi_\ell}\}_{\ell=0}^{n-1}$, which, together with the {\it a priori} known set of probe projectors, will be the inputs to an iterative phase retrieval algorithm designed to reconstruct $|\psi\rangle$. The entire process of quantum state ptychography is illustrated in Fig.~\ref{fig:ptico_clas}(c).

The iterative reconstruction algorithm we adopt is an adapted version of the PIE \cite{Rodenburg04,Faulkner05}. It proceeds by taking the estimate through the same steps that the ensemble of quantum systems was taken, imposing the measured data, and applying an update rule to the estimate. Its steps are the following:
\begin{enumerate}
\item Start with a random estimate of the input state: \begin{equation} |\phi\rangle=\sum_{k=0}^{d-1}\gamma_k|k\rangle . \end{equation}
\item Apply the $\ell$-th probe projector to $|\phi\rangle$: \begin{equation}\label{eq:psil} |\phi_{\ell}\rangle=\hat{P}_{\ell}|\phi\rangle = \sum_k \gamma_{k\ell} |k\rangle . \end{equation}
\item Apply the QFT to $|\phi_\ell\rangle$: \begin{equation}|\tilde{\phi}_{\ell}\rangle=\hat{F}_d |\phi_{\ell}\rangle = \sum_{k=0}^{d-1} \tilde{\gamma}_{k\ell} |k\rangle . \end{equation}
\item Use the  $\ell$-th measured ptychographic data, $\sqrt{\Pi_\ell}$, to correct the moduli of the coefficients of $|\tilde{\phi}_{\ell}\rangle$, keeping their phases: \begin{equation}|\tilde{\phi}_{\ell}'\rangle = \sqrt{\mathcal{N}} \sum_{k=0}^{d-1} |\tilde{c}_{k\ell}| e^{i\arg{\tilde{\gamma}}_{k\ell}} |k\rangle . \end{equation}
\item Apply the inverse QFT to obtain an updated estimate for the output state: \begin{equation}|\phi_{\ell}'\rangle=\hat{F}^{-1}_d |\tilde{\phi}_{\ell}'\rangle . \end{equation}
\item Update the current estimate of the input state: \begin{equation}\label{eq:update} |\phi'\rangle = |\phi\rangle + \beta\hat{P}_{\ell} \big(|\phi_{\ell}'\rangle - |\phi_\ell\rangle \big), \end{equation} where $\beta$ is a feedback parameter, roughly within $(0,2]$, that controls the step-size of the update and can be adjusted to improve convergence (see Methods).
\item Use this updated estimate as input to repeat the steps (ii)--(vi) with a new value of~$\ell$.
\end{enumerate}
This sequence is summarized in the diagram of Fig.~\ref{fig:ptico_clas}(d):  a single PIE iteration consists of  $n$ iterations through the closed loop [steps (ii)--(vii)], where each probe projector and corresponding ptychographic data is used once to update the state~estimate. At each iteration we calculate the relative distance between the current and updated estimates, i.e., $D=\||\phi'\rangle-|\phi\rangle\|^2/\||\phi\rangle\|^2$;  the algorithm terminates when it achieves either a sufficiently small value of $D$ or a preset maximum number of PIE iterations, delivering a pure state that must be normalized.

\subsection*{Equivalence to an alternating gradient search}

By following the same reasoning that has been applied in classical ptychography\cite{Sicairos08,Fienup82}, we will show that the iterative algorithm we propose is equivalent to an alternating gradient search with a weighting factor. More precisely, we will show that an iteration using the projector $\hat{P}_\ell$ is a gradient descent with respect to the error metric
\begin{equation}
  E_\ell = \sum_k \left[ |\tilde{\gamma}_{k\ell}| - |\tilde{c}_{k\ell}| \right]^2.
\end{equation}
We will proceed to calculate the gradient of the metrics $E_\ell$ with respect to the coefficients $\gamma_{k\ell}$ of the estimate [equation~(\ref{eq:psil})] and find the direction of fastest descent. To this end, we calculate the complex gradient of $E_\ell$,
\begin{equation}
  \left[ \frac{\partial}{\partial \gamma^{\rm R}_{j\ell}} + i \frac{\partial}{\partial \gamma^{\rm I}_{j\ell}} \right] E_\ell,
\end{equation}
where $\gamma^{\rm R}_{j\ell}$ and $\gamma^{\rm I}_{j\ell}$ denote the real and imaginary parts of $\gamma_{j\ell}$, respectively. The chain rule gives us
\begin{equation}\label{eq:dedg}
  \frac{\partial E_\ell}{\partial \gamma_{j\ell}} = 2\sum_k \left[ |\tilde{\gamma}_{k\ell}| - |\tilde{c}_{k\ell}| \right] \frac{\partial |\tilde{\gamma}_{k\ell}|}{\partial \gamma_{j\ell}},
\end{equation}
where $\partial/\partial \gamma_{j\ell}$ denote differentiation with respect to either the real or imaginary part of $\gamma_{j\ell}$.  The chain rule gives further
\begin{equation}\label{eq:dabsdg}
  \frac{\partial |\tilde{\gamma}_{k\ell}|}{\partial \gamma_{j\ell}} = \frac{1}{2|\tilde{\gamma}_{k\ell}|} \left\{ \tilde{\gamma}_{k\ell}\left[\frac{\partial \tilde{\gamma}_{k\ell}}{\partial \gamma_{j\ell}}\right]^* + \tilde{\gamma}^*_{k\ell}\frac{\partial \tilde{\gamma}_{k\ell}}{\partial \gamma_{j\ell}} \right\}.
\end{equation}
The derivatives of $\tilde{\gamma}_{k\ell}$ can be easily calculated through their Fourier transform relation to the coefficients $\gamma_{j\ell}$, giving us
\begin{equation}
  \frac{\partial \tilde{\gamma}_{k\ell}}{\partial \gamma^{\rm R}_{j\ell}} = -i\frac{\partial \tilde{\gamma}_{k\ell}}{\partial \gamma^{\rm I}_{j\ell}} = \frac{1}{\sqrt{d}} e^{-i2\pi jk/d}.
\end{equation}
Inserting these into equation~(\ref{eq:dabsdg}), we arrive at
\begin{equation}\label{eq:dabsdgr}
  \frac{\partial |\tilde{\gamma}_{k\ell}|}{\partial \gamma^{\theta}_{j\ell}} = \frac{i^{\delta_{\theta I}}}{\sqrt{d}} \frac{(-1)^{\delta_{\theta I}} \tilde{\gamma}_{k\ell}e^{i2\pi jk/d} + \tilde{\gamma}^*_{k\ell}e^{-i2\pi jk/d}}{2|\tilde{\gamma}_{k\ell}|},
\end{equation}
for $\theta=R,I$ and $\delta$ denoting a Kronecker delta. Plugging equation~(\ref{eq:dabsdgr}) back into equation~(\ref{eq:dedg}), we get
\begin{equation}
  \frac{\partial E_\ell}{\partial \gamma^{\rm \theta}_{j\ell}}  = (-i)^{\delta_{\theta I}} \left[ \left( \gamma_{j\ell} - \gamma'_{j\ell} \right) + (-1)^{\delta_{\theta I}}\left( \gamma^*_{j\ell} - \gamma'^*_{j\ell} \right) \right],
\end{equation}
where we used $\tilde{\gamma}'_{k\ell}=|\tilde{c}_{k\ell}|e^{i \arg \tilde{\gamma}_{k\ell}}$. Therefore, we have
\begin{equation}\label{eq:gdesc}
  \left[ \frac{\partial}{\partial \gamma^{\rm R}_{j\ell}} + i\frac{\partial}{\partial \gamma^{\rm I}_{j\ell}} \right] E_\ell = 2\left[ \gamma_{j\ell}-\gamma'_{j\ell} \right].
\end{equation}
This result should be compared to equation~(\ref{eq:update}). To make this more convenient, let us take the latter's inner product with $|j\rangle$: if we let $\hat{P}_\ell |j\rangle = \epsilon_{\ell j} |j\rangle$, with $\epsilon_{\ell j}=1$ in case $\hat{P}_\ell$ includes dimension $j$ and $\epsilon_{\ell j}=0$ otherwise, we get
\begin{equation}\label{eq:gdescorr}
  \gamma'_{j} = \gamma_j + \epsilon_{\ell j}\beta \left( \gamma_{j\ell}' - \gamma_{j\ell} \right).
\end{equation}
Therefore the update term $\epsilon_{\ell j}\beta( \gamma_{j\ell}' - \gamma_{j\ell})$ is a steepest descent step [equation~(\ref{eq:gdesc})] corrected by the factor $\beta\epsilon_{\ell j}$. While $\beta$ controls the size of the step, $\epsilon_{\ell j}$ only allows for correcting in the dimensions which have been addressed by $\hat{P}_\ell$. This is similar to a Wiener filter approach \cite{Rodenburg04}, where one trusts more on the correction term at points where the illumination was more intense and thus had a better signal-to-noise ratio. In our case, we have the analogue of a binary illumination, that is, each dimension either is or is not ``illuminated'', so that the Wiener-filter operator would just reduce to $\hat{P}_\ell$. Since the method uses several values of $\ell$, it is equivalent to an alternating gradient search.

The use of several different metrics is a mechanism for evading stagnation, which is a risk because of the algorithm's gradient-search nature. It is very unlikely that all the metrics have a local minimum at the same point. Therefore, in case the algorithm got stagnated at a local minimum of one of the metrics, an iteration corresponding to an $E_\ell$ which does not have a local minimum at that point would move the estimate away from it, and the algorithm would then proceed towards a common minimum.

The nature of the postprocessing algorithm demonstrated here is not, by itself, a guarantee of successful reconstructions. For this, it must be fed with ptychographic data having sufficient diversity and redundancy, which are determined by the defining features of the set of probe projectors (their form, $n$, $r$, and $\mathcal{O}$). In that case, we will provide strong numerical evidence that the algorithm succeeds in reconstructing an unknow pure state without ambiguity (see next section). Before that, we must specify the sets of projectors we shall work with.

\subsection*{Construction of probe projectors}
We have constructed three families of probe projectors to study the ptychographic protocol outlined above. We considered, initially, the projectors given by 
\begin{equation}   \label{eq:det_operator}
\hat{P}_\ell=\sum_{j=0}^{r-1}|j\oplus s_\ell\rangle\langle j\oplus s_\ell|,
\end{equation}
where $r$ is its rank, $\oplus$ denotes addition modulo $d$ and $s_\ell$ is a nonnegative integer that sets the skip between adjacent operators and  may be arranged in a $n$-entry vector $\mathbf{s}^{(n)}=(s_0,\ldots,s_{n-1})$. This is perhaps the simplest choice of projectors, with all of them being diagonal in the computational basis and encompassing contiguous dimensions of the Hilbert space.

\begin{figure}[tb]
  \centering
  \includegraphics[width=0.6\linewidth]{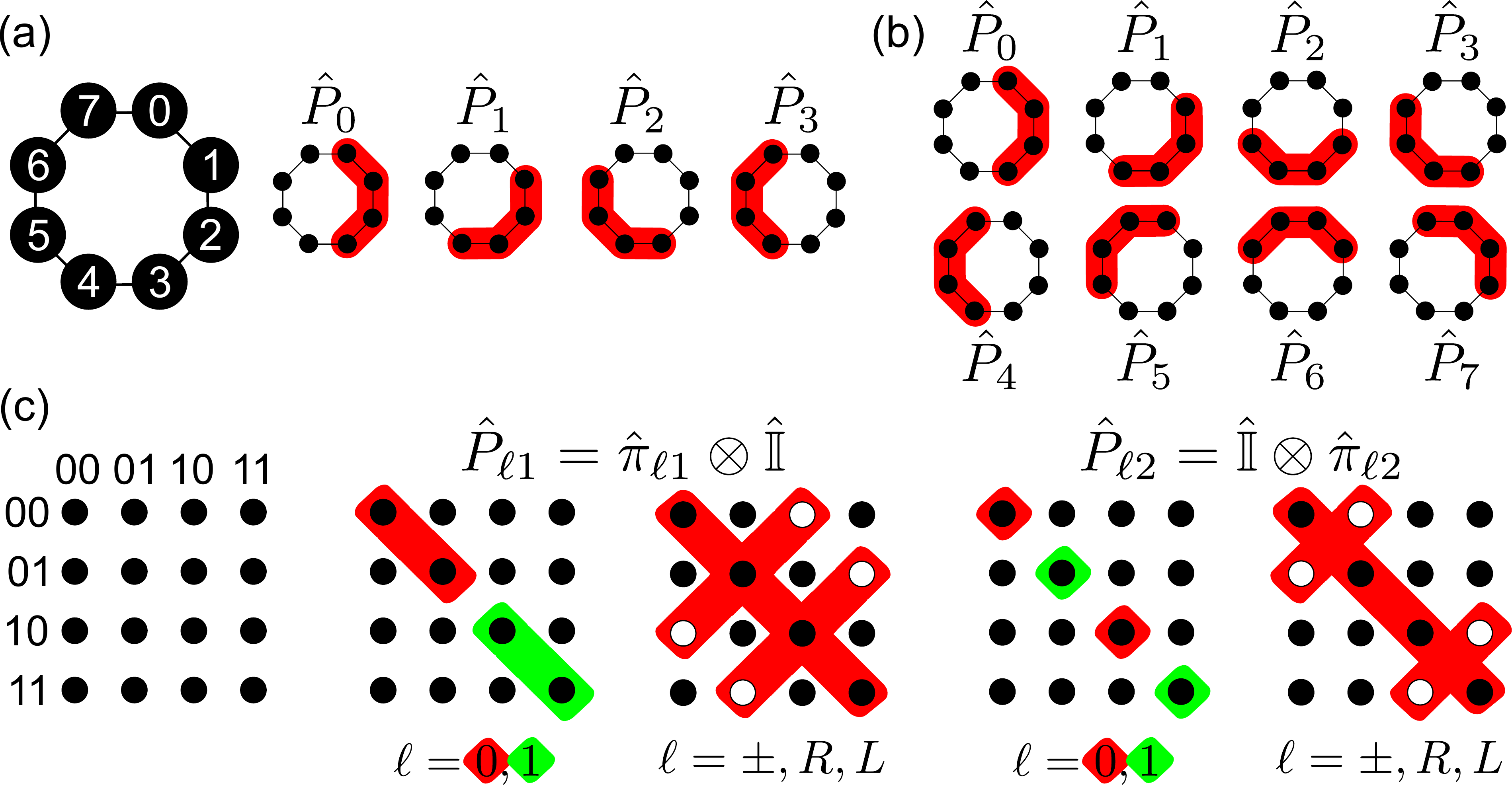}
  \caption{\textbf{Schematic representation of the probe projectors.} For the families in the form given by equation~(\ref{eq:det_operator}) and $d=8$ we have: (a) $n=4$ and (b) $n=8$ rank-4 projectors, acting as filters in the dimensions of $\mathcal{H}_d$. For the multiqubit family given by  equation~(\ref{eq:mult_operator}), (c) highlights their nonzero matrix components in the computational basis for a two-qubit system; the empty circles indicate the components multiplied by a phase factor that depends on $\ell$.  These projectors have rank 2.}
  \label{fig:projs}
\end{figure}

We derived two families of the form given above. In both cases, we used ranks around $d/2$, chosen from a numerical analysis seeking those that optimized PIE's convergence  (see Methods). For even dimensions we used $r=d/2$ whereas for the odd ones we alternated between $r=\lfloor d/2 \rfloor$ and $r=\lceil d/2 \rceil$, selecting the one which provided better reconstructions. In general, both ranks gave similar results. The first family used $n=4$ projectors about equally spaced along the dimensions of $\mathcal{H}_d$. This family requires $4d$ measurements to be carried, which is an amount comparable to other works in the literature of pure state reconstruction\cite{Goyeneche15,Carmeli16,Stefano17}. In this case, the vector of skips is given by
$ \mathbf{s}^{(4)}=\left( 0,\left\lceil\frac{d-r-2}{3}\right\rceil,2\left\lceil\frac{d-r-2}{3}\right\rceil,\left\lceil\frac{d}{2}\right\rceil\right), $
which will give an average overlap of $\mathcal{O}=2/3$.
The second family was made of $n=d$ operators with a skip vector $\mathbf{s}^{(d)}=(0,1,\ldots,d-1)$, which gives an overlap $\mathcal{O}=1-1/r$. This family is overcomplete, as it requires $d^2$ measurements to be carried, much more than is required for pure state reconstruction. However, it is capable of reconstructing a class of states that the first family is not, but which forms a zero-measure set in $\mathcal{H}_d$. We will leave this discussion to the next section, though. Figures~\ref{fig:projs}(a) and \ref{fig:projs}(b) illustrate the action of these families of probe projectors on the input states considering $d=8$.

The third and last family was conceived for the case where the quantum system is comprised of $N$ qubits. In this case, it is desirable to avoid operations that involve more than one qubit at a time,  and use only local operations instead. This is because the latter are easier to implement experimentally \cite{Klimov08}. We considered, therefore, the set of $n=6N$ probe projectors given by
\begin{equation}\label{eq:mult_operator}
  \hat{P}_{\ell j} = \hat{\pi}_{\ell j}\otimes\hat{\mathbb{I}}^{\otimes N-1},
\end{equation}
where $\hat{\pi}_{\ell j}=|\ell_j\rangle\langle\ell_j|$ are projectors onto the eigenstates of the Pauli operators $\hat{\sigma}_x$ ($\ell=+,-$), $\hat{\sigma}_y$ ($\ell=R,L$) and $\hat{\sigma}_z$ ($\ell=0,1$) of the $j$-th qubit, and $\hat{\mathbb{I}}$ is the identity in the qubit space. Therefore, $\hat{P}_{\ell j}$ projects the part of the state on the $j$-th qubit subspace while leaving the remainder unchanged. These probe projectors have rank $r=2^{N-1}$ and an overlap $\mathcal{O} = 1/2+\delta_{jj'}(|\langle\ell|\ell'\rangle|^2-1/2)$; most of them ($\ell\neq0,1$), unlike those in equation~(\ref{eq:det_operator}), are not diagonal in the computational basis. Thus, they are better visualized by their matrix components in that basis. Figure~\ref{fig:projs}(c) illustrates this for a two-qubit system.

It is worth noting that there is plenty of freedom in constructing the family of projectors. The only requirements are those established in the beginning of this section, namely, that they must have some overlap and address all levels in $\mathcal{H}_d$. The reader is encouraged to pursue other families that might be best suited for his/her experimental setup, for example.

\section*{Results} \label{sec:numresults}

\subsection*{Blind reconstruction}
\begin{figure*}[tb]
  \centering
  \includegraphics[width=0.7\textwidth]{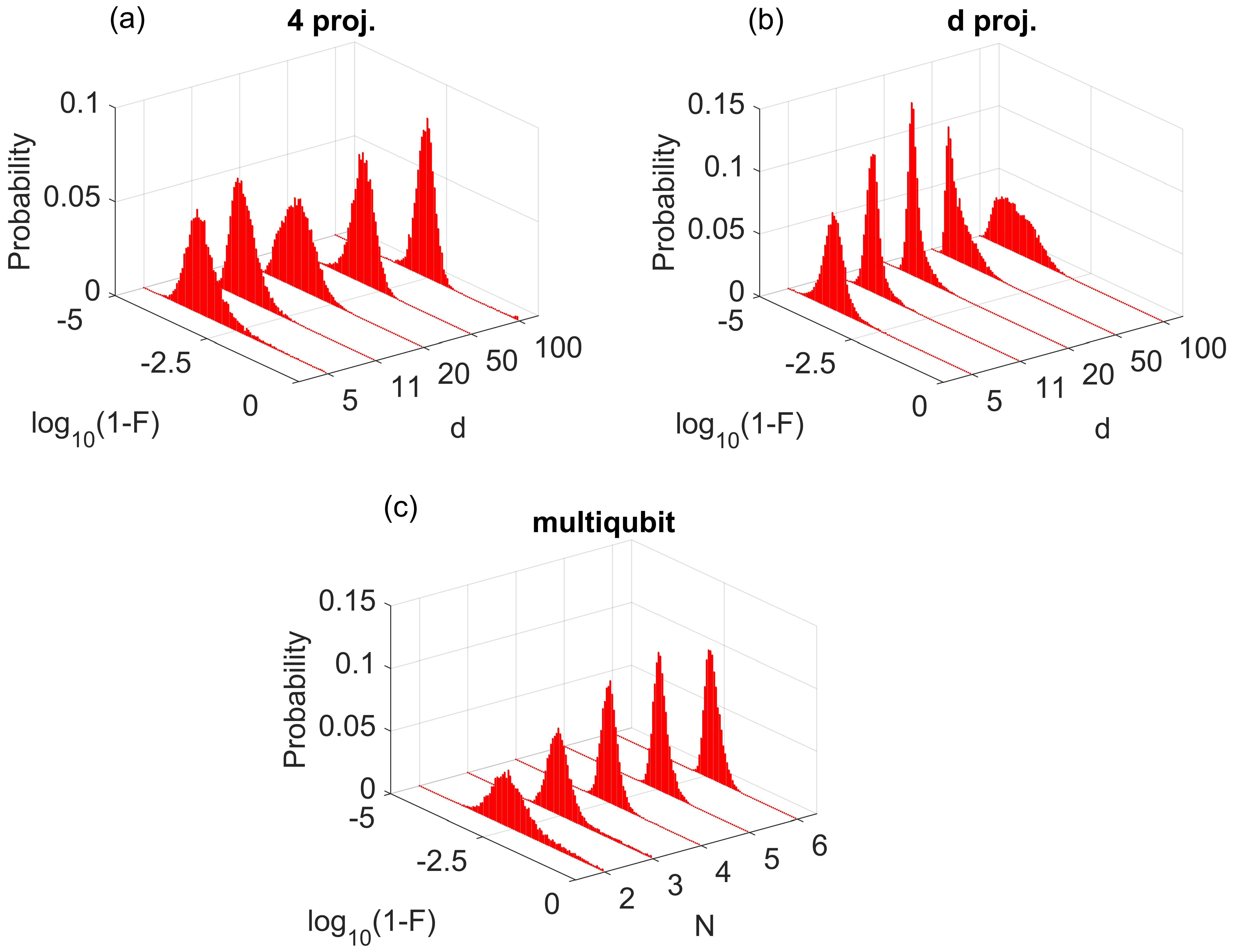}
  \caption{\textbf{Blind reconstruction results.} Histograms of infidelities for reconstructions using the (a) $n=4$, (b) $n=d$ and (c) multiqubit families of projectors. We used ranks $r=3,6,10,25,50$ for dimensions $d=5,11,20,50,100$, respectively.} 
 \label{fig:fids_noise}
\end{figure*}

Using the three sets of probe projectors defined above, we performed a numerical study of the quantum ptychographic protocol. Our study comprised several dimensions up to $d=100$. For each $d$, $10^4$ input quantum states were randomly generated according to the Haar measure. The ptychographic data sets corresponding to these states were used to simulate experimental detections (more details next), which were then fed into the PIE algorithm. We chose for our feedback parameter the value $\beta=1.5$, optimized numerically (see Methods). Our stop criteria consisted of two clauses: reaching $D<10^{-5}$ or 100 PIE iterations, whichever happened first. If the algorithm was stopped by the second clause, we made it start again with a new random estimate. We allowed up to 100 such reinitializations. At the end, the quality of the reconstruction is quantified by computing the fidelity $F=|\langle\phi_{\rm \textsc{pie}}|\psi\rangle|^2$ and infidelity $I=1-F$ between the input state, $|\psi\rangle$, and the normalized estimate of the algorithm, $|\phi_{\rm \textsc{pie}}\rangle$.

We simulated the imperfect generation of states and the random nature of detections, which are unavoidable sources of uncertainty in actual experiments, with depolarization and Poissonian noise, respectively. The first can be modeled as a random fluctuation in the density matrix of the pure state ($|\psi\rangle\langle\psi|$), so that the generated state will be
\begin{equation}
  \hat{\rho}=(1-\eta)|\psi\rangle\langle\psi|+\eta\hat{\rho}_{\rm rand},
\end{equation}
where $\eta$ is the noise level, and the random perturbation $\hat{\rho}_{\rm rand}$ is drawn according to the Hilbert-Schmidt measure in the mixed-state space\cite{Zyczkowski11}. The probe projection followed by the QFT will produce $\hat{\rho}_\ell=\hat{P}_\ell \hat{\rho}\hat{P}_\ell$ and $\hat{\sigma}_\ell=\hat{F}_d \hat{\rho}_{\ell}\hat{F}_d^{-1}$. The diagonal components of $\hat{\sigma}_\ell$, denoted by $\{|\widetilde{C}_{k\ell}|^2\}_{k=0}^{d-1}$, provided the simulated data to which we applied a Poisson distribution of average $\lambda|\widetilde{C}_{k\ell}|^2$, where $\lambda$ is a count rate factor. In our simulations we used $\eta=0.05$ and $\lambda=10^3$, in agreement with values found in the literature \cite{Gross10,Kaznady09}, which we also checked to be realistic (see Methods). Note that although the noise will introduce mixedness in the input states, the PIE algorithm will treat the noisy data as if they came from a pure state, and it will deliver a pure state as well.

\begin{figure}[tb]
  \centering
  \includegraphics[width=0.7\linewidth]{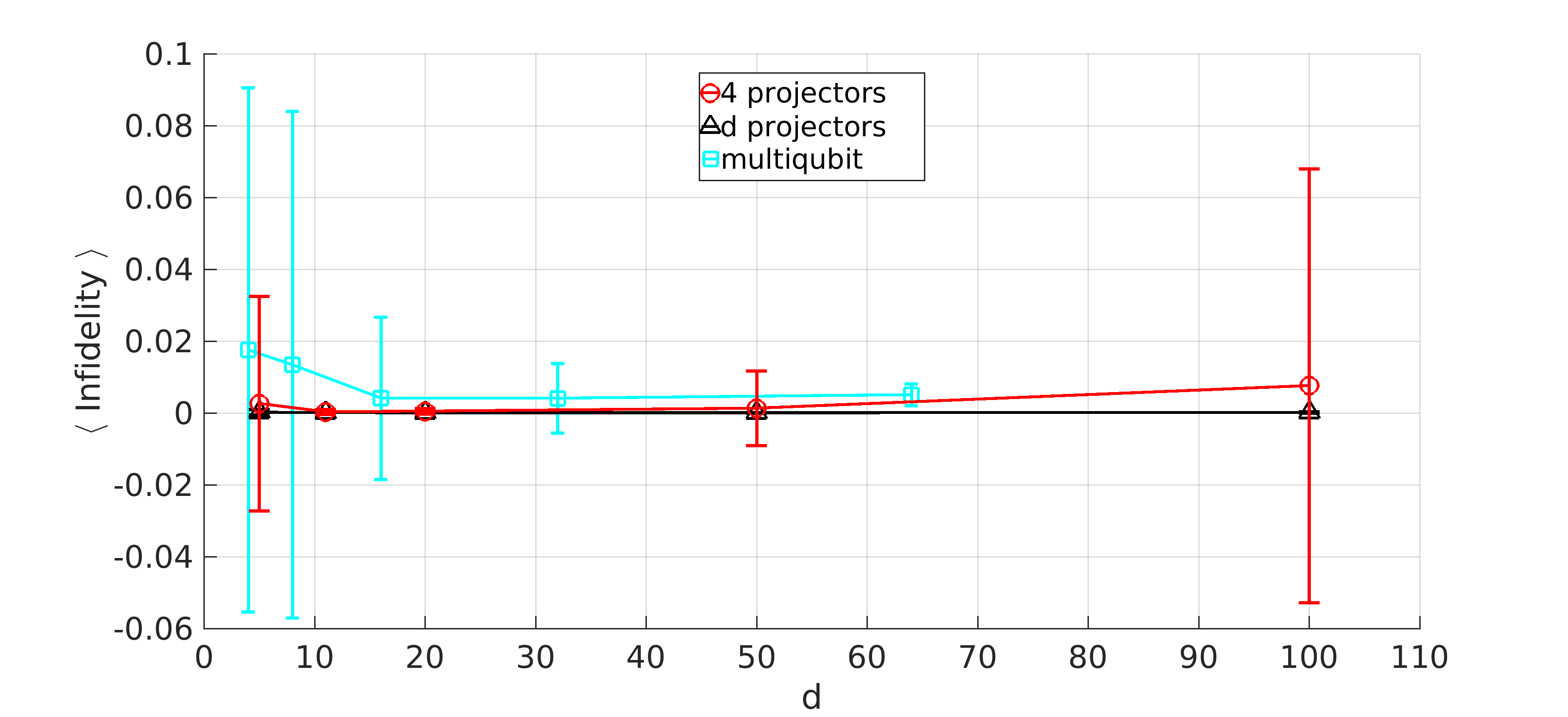}
  \caption{\textbf{Average infidelities of reconstructions}. Average infidelities of the reconstructions shown in Fig.~\ref{fig:fids_noise}, using the $n=4$ (red circles), $n=d$ (black triangles) and multiqubit (cyan squares) families of projectors. The standard deviations are shown as error bars. Overall, the infidelities were very close to zero as desired, especially for the $d$-projector family.}
  \label{fig:nits}
\end{figure}

Figure~\ref{fig:fids_noise} shows the histograms of infidelity obtained in this case and figure~\ref{fig:nits} shows their averages; overall, $I<10^{-2}$, which attests that the reconstructions were excellent in general. Only a small fraction of them were not satisfactory. In $d=100$, for instance, about 4\% of input states have not been well characterized (i.e., had fidelities below $0.9$) by the family with 4 projectors. A straightforward way to overcome this issue is including one or a few more probe projector(s) in the measurement at the expense of increasing the experimental cost. Alternatively, we may adapt to our protocol the recent improvements in the PIE algorithm that successfully handle difficult data sets \cite{Maiden17}, but this is beyond the scope of the present work.

We have observed that the $d$-projector family of projectors took reasonably longer times than the $4$-projector family to reconstruct the states. For example, to reconstruct $10^4$ states of dimension $d=11$ with projectors of rank $r=6$, the scheme with $n=4$ projectors took about $800$~s, while about $10 000$~s were needed for $n=d$. For $d=100$ and $r=50$, the reconstruction times went to about $40$~min and $6$~h for $n=4$ and $n=d$, respectively. This suggests that a greater diversity in the data makes the algorithm able to detect more imperfections in the estimate, so that more iterations are needed to reach the threshold fixed for $D$. To make an analogy, picture a carpenter that refines the shape of a wooden piece; by looking at the piece from two orthogonal angles, she is able to correct its shape to some extent. Then, by looking from an angle intermediate to those two, she is able to perceive other imperfections that were not evident and correct them. Therefore, by using more diverse data, the algorithm is able to correct the estimate in a manner that would not be possible by just using more iterations of less diverse data. On the other hand, it could also make more efficient corrections to the estimate, as they will now happen on more subspaces of $\mathcal{H}_d$, so that it is not trivial to foresee how data diversity impacts convergence.

The algorithm required a few thousands of iterations to converge, in general. However, each iteration is extremely fast since the estimate is just a complex vector (as opposed to an array, for example), so the operations that need to be carried out are comparatively simple. Overall, with a modest laptop, the reconstruction of a single state took from fractions of a second, at smaller dimensions, up to a few seconds at greater ones, as can be inferred from the reconstruction times mentioned above.

\subsection*{Non-overlapping projectors and sparse states}

In order to highlight the role of the overlap between probe projectors in the ptychographic measurements, we have also simulated reconstructions using non-overlapping $\hat{P}_\ell$'s. For $d=20$, $n=4$, $r=5$,  $\mathbf{s}^{(4)}=(0,5,10,15)$ in equation~(\ref{eq:det_operator}), and $10^4$ random states, we obtained fidelities ranging from $10^{-4}$ to $0.81$ with an average of $0.15$. These almost random results show that the multiple overlaps are crucial in the protocol: without them, the ptychographic problem becomes several disjoint standard phase retrieval problems \cite{Gerchberg72}, which are known to suffer from nonuniqueness and stagnation issues \cite{Sicairos08}.

The ptychographic method, with the three projector families used so far, will suffer difficulties in the reconstruction of sparse pure states, i.e., states in which most of the components are zero. It succeeds only if every nonzero component was addressed with at least one other nonzero component by some probe projector, so that they could interfere. This is hindered, in general, when the state is sparse. The results is that the algorithm falls into a disjoint group of phase retrieval problems, as discussed above. One way to avoid this is to use the already considered set of $n=d$ probe projectors [equation~(\ref{eq:det_operator}) with $\mathbf{s}^{(d)}=(0,1,\ldots,d-1)$], but now with rank $r\geq\lfloor d/2\rfloor+1$, because when the levels of $\mathcal{H}_d$ are addressed cyclically, the biggest distance between nonzero components will be $\lfloor d/2\rfloor$. A second way is to use an adaptive approach: first, one measures in the computational basis; if the state is verified to be sparse, then one applies the ptychographic method building the probe projectors according to the distribution of its nonzero components. Such an extra step is, in general, easy to carry out.

\section*{Discussion}

\noindent The ptychographic method introduced here requires a total number of $\mathcal{M}=nd$ measurement outcomes---$d$ QFT state-amplitudes for each of the $n$ probe projections. In this regard, its experimental cost will be determined by the number of $\hat{P}_{\ell}$'s adopted. Along with the specific form of the projectors, this number also defines the diversity of the ptychographic data set and its degree of redundancy  arising from the partially overlapping projections. When choosing $n$, one should be aware that a high value, although experimentally more demanding, provides more diversity and redundancy in the data set. This has consequences on the quality of the reconstructions and on the convergence of the algorithm, and our results revealed that there can be appreciable differences between projector families. Nevertheless, all the three families we studied gave good results, in spite of their differences, even with substantial amounts of noise included in the simulations.

\begin{figure}[tb]
  \centerline{\includegraphics[width=0.6\columnwidth]{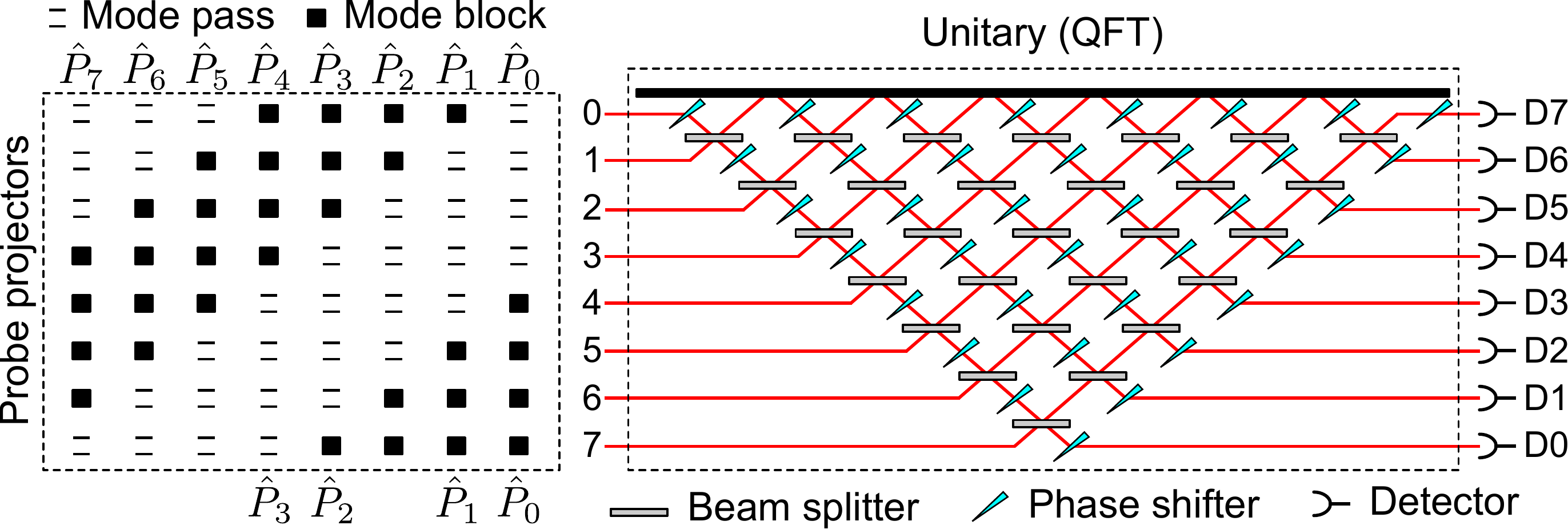}}
	\caption{\textbf{Scheme for ptychographic reconstruction of 8-dimensional quantum states in a multiport interferometer}. The rank-4 probe projectors given in equations~(\ref{eq:det_operator}) are implemented via mode filters: (top) $n=8$; (bottom) $n=4$.} 
	\label{fig:multiport}
\end{figure}

To illustrate the simplicity of the quantum ptychographic scheme and, at the same time, discuss other of its general aspects, let us consider $d$-dimensional states encoded in the propagation modes of single photons (or any other type of radiation). A multiport interferometer (MI), sketched in the right box of Fig.~\ref{fig:multiport} for $d=8$, can implement any unitary transformation on this encoding \cite{Reck94}. Under these circumstances, the probe projectors given by equation~(\ref{eq:det_operator}) would be realized by mode filters at the input ports of the interferometer, as shown in the left box of Fig.~\ref{fig:multiport}. By setting the MI to perform $\hat{F_d}$, the ptychographic data would be collected simply by shifting the mode filters $n$ times at the input ports and recording the counts at the output ports. For comparison, to reconstruct these states by measuring four or five observables \cite{Goyeneche08,Goyeneche15,Stefano17,Carmeli16}, the mode filters would not be necessary, but one would have to reconfigure the whole MI for each measurement basis employed. This shows a nice feature of the ptychographic method: the measurements are effectively performed in a single basis while the probe projectors are ``shifted'' through the Hilbert space.

From the experimental point of view, the implementation of quantum state ptychography should be straightforward. On one hand, the QFT has been experimentally realized in many different platforms including trapped ions\cite{Chiaverini05,Schindler13}, superconducting qubits\cite{Mariantoni11}, nuclear magnetic resonance\cite{Weinstein01}, neutral molecules\cite{Hosaka10}, and photonic systems\cite{Lu07,Lanyon07,Crespi16,Malik14,Prosser17}, both for single-particle and multiqubit scenarios. On the other hand, the families of probe projectors proposed here and given by equations~(\ref{eq:det_operator}) and (\ref{eq:mult_operator}) are both of simple implementation for these scenarios.

\section*{Conclusion}
We have proposed and numerically studied a method to reconstruct pure quantum states based on a powerful coherent diffractive imaging technique called ptychography \cite{Hoppe69,Rodenburg08,Faulkner04}. Our quantum ptychographic protocol was shown to be simple in regard to the required set of measurements and to possess a fast and robust postprocessing, regardless of the state-space dimension. It was also shown to be equivalent to an alternating-gradient search algorithm. Successful reconstructions were obtained in realistic noisy scenarios, which makes the method amenable to future experiments and a concrete alternative to standard tomographic techniques \cite{Goyeneche08,Goyeneche15,Stefano17,Carmeli16}.

Since the emergence of ptychography in its modern form with the works of Faulkner and Rodenburg \cite{Faulkner04,Rodenburg04}, the technique has evolved impressively. Subsequent advances included, in special, the recovery of the illumination probe \cite{Thibault09}, the use of other propagators rather than the Fourier transform \cite{Stockmar13}, and the handling of mixedness both in the probe and in the object \cite{Thibault13}. Our method, based on the simplest form of ptychography \cite{Faulkner04}, may follow a similar route and be extended in many directions, including the utilization of different types of probe operators and different measurement bases, the reconstruction of mixed states and processes both in discrete and continuous domains, among others.  Thus, quantum ptychography has the potential to become a valuable tool for quantum information science.

\section*{Methods} \label{sec:methods}

\subsection*{Optimization of the feedback parameter $\beta$}

The parameter $\beta$ [equation~(\ref{eq:update})] controls the step-size of the update in the PIE algorithm and can be adjusted to improve its convergence. For $\beta=1$, the algorithm corrects the estimate strictly in the subspace spanned by $\hat{P}_\ell$; higher values can make it progress faster and converge in less iterations; lower values can make it slower but more stable. Therefore, it is advisable to run a few reconstructions with several values of $\beta$ and compare their performance.

Using the same initial estimated and target state, we obtained the optimal $\beta$ by running the PIE and recording the relative distance $D=\||\phi'\rangle-|\phi\rangle\|^2/\||\phi\rangle\|^2$ between current and last estimate, for several values of the parameter. Figure~\ref{fig:betas} (left panel) shows the evolutions for a few values of $\beta$. The best progression was achieved by $\beta=1.5$, which we used in all later studies.

\subsection*{Optimal rank of projectors}
To determine the optimal $r$ of the probe projectors given by equation~(\ref{eq:det_operator}), we studied the convergence of the PIE algorithm as a function of this rank for a few state-space dimensions ($d=10,15,20$). For each combination of $d$ and $r$, we reconstructed $10^4$ random states and calculated the average number of iterations necessary until convergence. The results are shown in Fig.~\ref{fig:betas} (center panel) and indicate that a rank around $d/2$ works best. In our simulations we alternated between $\lfloor d/2 \rfloor$ and $\lceil d/2 \rceil$. In general, we verified that both, as well as any other close value, produced similar results regarding the quality of the reconstructions.

\subsection*{Noise simulation}
\begin{figure*}[tb]
  \centerline{
    \includegraphics[width=0.99\textwidth]{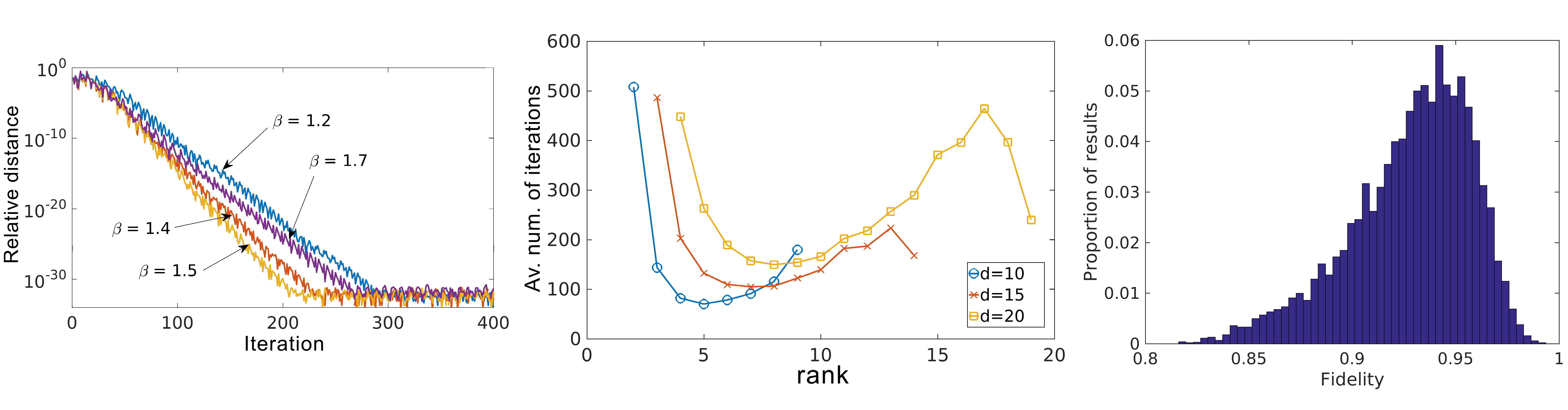}
  }
  
    \caption{\textbf{Adjustment of numerical parameters.} (Left panel) Evolution of the relative distance in the state estimate for different values of the feedback parameter, $\beta$. We found $\beta=1.5$ to achieve the fastest decrease, meaning that the PIE algorithm converged at a higher rate. (Center panel) Average number o iterations until convergence of the PIE algorithm as a function of the probe projector rank. Ranks around $d/2$ made the algorithm converge faster. (Right panel) Histogram of degraded fidelities; upon comparison with experimental works in the literature, it confirms that the noise parameters in our simulations were indeed realistic.}
	\label{fig:betas}
\end{figure*}

We introduced depolarization and Poissonian noise in the ptychographic data to study the protocol in a realistic scenario. As mentioned in the main text, we based our noise levels on experiments found in the literature, but we still wanted to verify if they were indeed realistic. To this end, we picked $10^4$ random pure states, degraded their amplitudes with the two kinds of noise and computed the fidelities with respect to the original states. Figure~\ref{fig:betas} (right panel) shows a histogram of the degraded fidelities, which are indeed comparable---and even lower---to actual experiments \cite{Varga14}.

\section*{Data availability}
The simulation codes and  datasets generated during the current study are available from the corresponding authors on reasonable request.


\section*{Acknowledgements}
This work was supported by FAPEMIG (APQ-00240-15), CNPq (407624/2018-0) and CNPq INCT-IQ (465469/2014-0). M.\ F.\ F. acknowledges financial support from CNPq (140359/2017-6.).

\section*{Author Contributions}
L.N. conceived the study. Both authors formulated the method. M.F.F. developed the reconstruction algorithm, demonstrated its gradient-search equivalence, performed the simulations and analyzed the results. Both authors wrote and revised the manuscript.



\end{document}